\documentclass[a4paper,12pt]{article}
\usepackage[
        left=3cm,
        right=2cm,
        top=3cm,
        bottom=2cm,
]{geometry}
\usepackage{lipsum}
\usepackage[T1]{fontenc}
\usepackage[utf8]{inputenc}
\usepackage{amssymb}
\usepackage{amsmath}
\usepackage{amsfonts,relsize}
\usepackage{authblk}
\usepackage{graphicx}
\usepackage{hyperref}
\usepackage{lineno}
\usepackage{cite}
\usepackage{setspace}
\usepackage{doi,verbatim}
\usepackage{xcolor}
\begin{document}
%
%\font\myfont=cmr12 at 10pt
%
%\title{\myfont \textbf{2D Langevin dynamics with white and correlated noises} }
%
\title{ \textbf{Memory and irreversibility on two-dimensional overdamped Brownian dynamics} }
\author[1]{Eduardo dos S. Nascimento\thanks{edusantos18@esp.puc-rio.br}}
\author[1,2]{Welles A. M. Morgado\thanks{welles@puc-rio.br} }
\affil[1]{Dept. of Physics, PUC-Rio, Rio de Janeiro, Brazil}
\affil[2]{National Institute for Science and Technology - Complex Systems, Brazil}

\date{} 
%
%
%
%\twocolumn[
%  \begin{@twocolumnfalse}
\maketitle
\begin{abstract}
We consider the effects of memory on the stationary behavior of a two-dimensional Langevin dynamics in a confining potential. 
The system is treated in an overdamped approximation and the degrees of freedom are under the influence of distinct 
kinds of stochastic forces, described by Gaussian white and colored noises, as well as different effective temperatures. 
The joint distribution function is calculated by time-averaging approaches, and the long-term behavior is analyzed. 
We determine the influence of noise temporal correlations on the steady-state behavior of heat flux and entropy production. 
Non-Markovian effects lead to a decaying heat exchange with spring force parameter, which is in contrast to the usual linear 
dependence when only Gaussian white noises are presented in overdamped treatments. Also, the model exhibits non-equilibrium states 
characterized by a decreasing entropy production with memory time-scale. 
\end{abstract}
{\bf Keywords:} Non-equilibrium; stochastic thermodynamics; memory effects; Langevin dynamics
%
%\vspace{2cm}
%
%
% \end{@twocolumnfalse}
%]
%
%

%
%
\section{Introduction} \label{Intro}
%

%Statistical mechanics is a powerfull theory of morden physical sciences that allow the 
%understanding of interactiong many-body systems, and its different formulations lead 
%to general results and methods with applications far beyond mascroscopic thermodynamics 
%of equilibrium originaly associated with. 

The current interest in emergent properties and thermodynamics of mesoscopic and small systems 
give rise to very rich discussions and investigations about the fundamental concepts and applications of statistical physics in non-equilibrium  \cite{Gallavotti2004,BroeckMeurs2004,Seifert2005,MarconiVulpiani2008,Seifert2012,
CelaniAurell2012,PUGLISI20171,BerutCiliberto2016A,CilibertoTanase2013}. 
Many of these studies can be addressed by means of Langevin dynamics (LD) \cite{Kubo1966,Zwanzig,Gardiner,
Sekimoto2010,CilibertoTanase2013,FogedbyImparato2011,CrisantiVillamaina2012,BerutCiliberto2016,BerutCiliberto2016A}, 
which emphasizes the role of distinct time-scale contributions to the temporal evolution of many-particle systems 
described in terms of effective degrees of freedom. Despite the simplicity, LD provide theoretical framework for  modelling stochastic properties that characterize different kinds of complex systems in physics, chemistry and biology \cite{Gardiner}. Also, by means of LD, it is possible to develop stochastic analogs of thermodynamic quantities such as heat and work that may contribute for the understanding of non-equilibrium systems 
\cite{Sekimoto1998,Sekimoto2010,MurashitaEsposito2016}.

Paradigmatic models for studying non-equilibrium behavior are usually formulated in terms of Langevin equations with many different kinds of stochastic forces, usually described by Gaussian 
\cite{SoaresMorgado2008,CrisantiVillamaina2012,DotsenkoOshanin2013,MancoisWilkowski2018} 
and non-Gaussian noises \cite{KanazawaHayakawa2013,MedeirosQueiros2015,Queiros2016,MorgadoQueiroz2016}. 
The simplest case of an overdamped two-dimensional system in symmetric harmonic potential, 
where each degree of freedom is associated with a different thermal bath, as discussed by Dotsenko and 
collaborators \cite{DotsenkoOshanin2013}, present a non-equilibrium distribution which leads to steady-states 
that exhibit probability currents that varies spatially in a nontrivial fashion. Similar results are found if one 
consider inertial effects as well asymmetric potentials, according to calculations and simulations developed by 
Mancois and collaborators \cite{MancoisWilkowski2018}. The emergency of stationary states with spatial-dependent 
probability flux, which give rise to non-zero mean angular velocity for Brownian particles, is due to the interplay between different temperatures and coupled degrees of freedom.

Another way to obtain stationary behavior of non-equilibrium in Langevin systems is through 
stochastic forces with memory, or temporal correlations, which may provide violations of 
fluctuation-dissipation relations \cite{Kubo1966,Zwanzig,PuglisiVillamaina2009,VillamainaVulpiani2009}. An investigation developed by Puglisi and Villamaina \cite{PuglisiVillamaina2009} has shown that, for a massive one-dimensional Langevin system with many colored noises, memory affects irreversibility through contributions that behave as effective non-conservative forces. Also, the analysis of Villamaina and collaborators \cite{VillamainaVulpiani2009} discusses the consequences of adopting a reasonable set of degrees of freedom, and related fluctuation-dissipation relations, in order to characterize the stationary states of a Brownian particle with memory.
In fact, the inclusion of time-correlated Langevin forces affects some dynamical aspects of Brownian motion, specially when 
inertial contributions are not properly considered. According to investigations of Nascimento  and Morgado \cite{NascimentoMorgado2019}, an overdamped Brownian particle with memory evolves to a non-equilibrium distribution in one dimension. Notice that the usual dissipation memory kernel is related to colored noise second cumulant in order to promote the correct Boltzmann-Gibbs (BG) 
statistics for long-term behavior of a massive system \cite{Kubo1966,Zwanzig}. Although equilibrium is not 
achieved by considering time-correlated noise in overdamped treatments, the inclusion of an additional weak white noise may regularize stationary states and BG is recovered  \cite{NascimentoMorgado2019}. 

Also, disregarding "mass" effects may provide artifact results for heat exchanges in many-bath environments. For a model consisting of coupled,  two-temperature, overdamped Langevin equations that describe thermal conduction through degrees of freedom interacting via harmonic forces, Sekimoto \cite{Sekimoto2010} has show that heat flux may present a divergence behavior as the spring force constant $k \to \infty$. However, this nonphysical result is avoided if one considers inertial contributions. For a Brownian particle under the influence of many thermal baths, which also exhibits a divergent heat flux \cite{ParrondoEspanol1996}, Murashita and Esposito \cite{MurashitaEsposito2016} develop extensive calculations in order to properly consider the  stochastic thermodynamics in overdamped conditions. In fact, there exist relevant contributions to heat conduction that come from dynamical evolution of momentum variables. Overdamped treatments also affect results associated with entropy production in systems that presents temperature gradients, which leads to a kind of entropy anomaly with vanishing inertia effects, as discussed by Celani and collaborators  \cite{CelaniAurell2012}.

Overdamped approximation models are important techniques which simplify the analysis of LD and, for some cases, give reasonable physical insights, but the absence of inertia should be considered with care on the statistical behavior of Non-Markovian systems. Then, in order to investigate the interplay between memory and irreversibility on overdamped situations, we revisit the problem of two-dimensional Brownian motion in contact with two thermal baths at different temperatures. We consider massless coupled Langevin equations, in a harmonic inter-particle force field, under the influence of Gaussian white and colored noises, each one acting on a distinct degree of freedom, and a dissipation memory kernel. The stationary probability distribution is calculated  by using time-averaging approaches, and non-Gibbsian and Gibbsian states can be identified, depending on model parameters. We determine the stochastic thermodynamics of the system, which lead to a  memory-dependent heat flux that decays with  spring constant force. This is very different from the usual linear dependence behavior exhibited by the case with two Gaussian white noises and indicates that temporal correlations affects the heat conduction for overdamped treatments in a nontrivial way. Also, we show that the memory kernel time-scale contributes to decreasing the entropy production for steady-states. Finally, we show that, for finite time correlations, one can find a stationary behavior of non-equilibrium that presents a zero entropy generation even when bath temperatures are equal. 

In this work, we emphasize only memory effects on massless Brownian dynamics in two-dimensional harmonic trap, which Markovian limit exhibits reasonable qualitatively physical results for finite values of spring force constant. Massive cases with colored noises usually present a very complicated mathematical structure to deal with analytically, even for one-dimensional cases \cite{SoaresMorgado2008}.

The paper is organized as follows. In Sect.\ref{2DLangevin}, we define the model of a two-dimensional LD in a overdamped approximation. We calculate the probability distribution and the stationary behavior in Sect.\ref{StationaryProb}. The heat flux and entropy production is determined in Sect.\ref{HeatEntropy} for steady-states. The conclusions are presented in Sect.\ref{Conclusions}.

\section{Langevin system in an harmonic potential} \label{2DLangevin}
\begin{figure}
 \centering
 \includegraphics[scale=0.4]{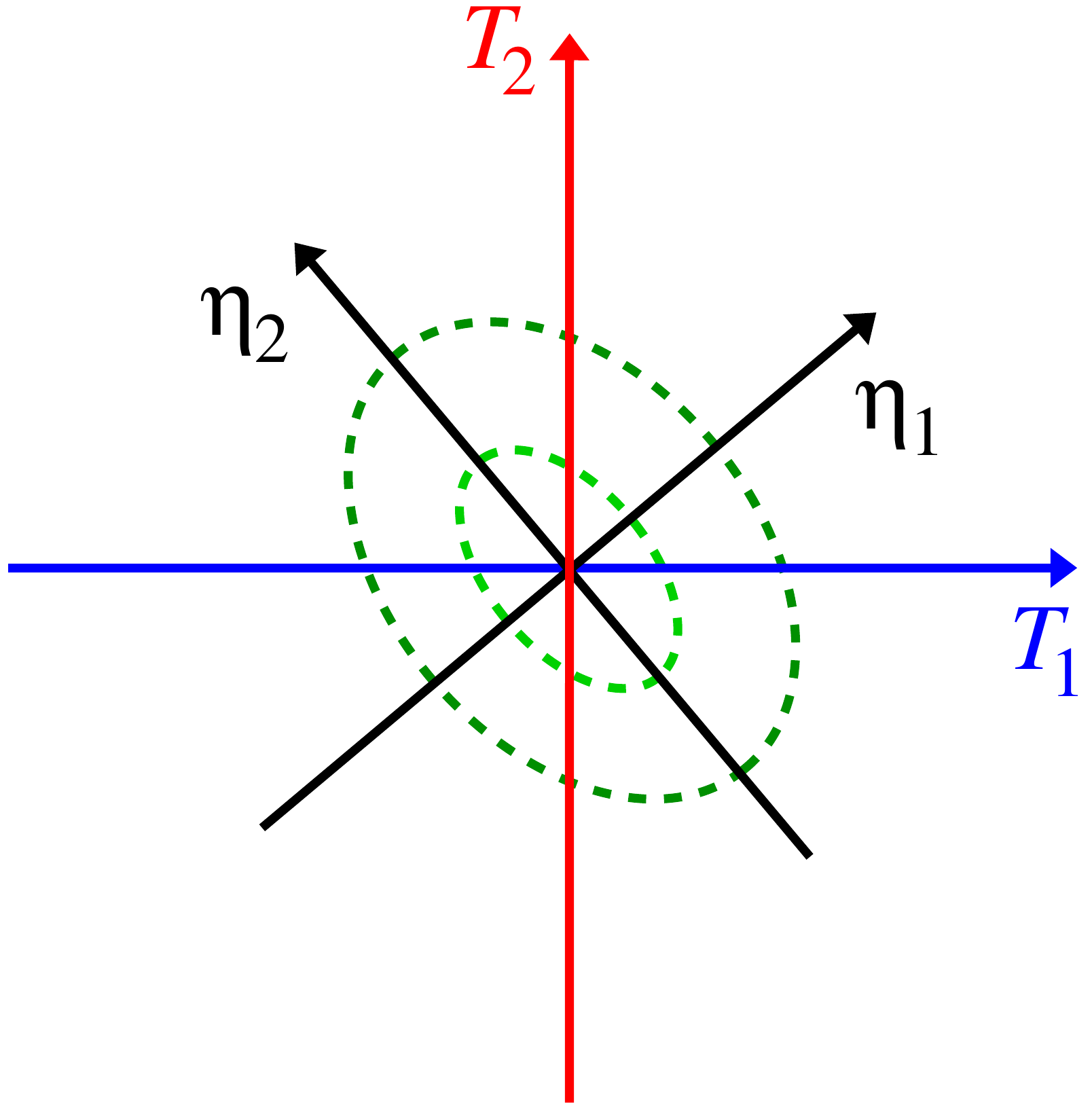}
 \caption{ Principal axes for a two-dimensional Brownian particle. 
  Stochastic forces act on ``temperature axes'' $T_1$ and $T_2$.
  The eigenframe of the harmonic potential is characterized by $\eta_1$ and $\eta_2$. 
  For different values of potential parameters $k$ and $u$, it is possible to 
  identify distinct equipotential  curves (green dashed lines).}
\label{2DLan}
\end{figure}

We consider a Brownian particle moving in two dimensions, with degrees of freedom $x_1$ and $x_2$, 
under the influence of a quadratic potential,
\begin{equation} \label{Potential}
 \begin{split}
  U(\textbf{x}) &= U(x_1,x_2) = \frac{k}{2}\left( x_1^{2} + x_2^{2} + 
  2\,u\,x_1\,x_2 \right), \\
               &= \textbf{x} \cdot \mathbb{V} \cdot \textbf{x},
 \end{split}    
\end{equation} 
where 
\begin{equation*}
 \textbf{x} = \begin{pmatrix}
  x_1 \\
  x_2
 \end{pmatrix} \quad
\textrm{and} \quad 
 \mathbb{V} = \frac{k}{2}\begin{pmatrix}
  1 & u \\
  u & 1
 \end{pmatrix}.
\end{equation*}
%
%
\begin{comment}
\begin{equation}
 \textbf{x} = \begin{pmatrix}
  x_1 \\
  x_2
 \end{pmatrix},
\end{equation}
%
and
%
\begin{equation}
 \mathbb{V} = \frac{k}{2}\begin{pmatrix}
  1 & u \\
  u & 1
 \end{pmatrix}.
\end{equation}
%
\end{comment}
%
Notice that we should consider $u^2<1$ in order to assure the confining aspect 
of the potential. Each degree of freedom is coupled to a thermal bath, one described by white noise and 
other represented by a colored noise, respectively, both of Gaussian character. Also, we assume baths at 
different ``temperatures'' $T_1$ and $T_2$. On heuristic grounds, one can think of two 
Langevin forces acting along the ``temperature axes'', which coincides with the Cartesian 
frame $x_1$ and $x_2$, as well the eigenframe associated with harmonic potential, see 
Fig.\ref{2DLan}. For nonzero value of coupling parameter $u$, the directions of 
stochastic forces do not coincide with the principal axes of the quadratic 
form \eqref{Potential}. 
%
%As a result, nontrivial energetic fluxes may be stabelished 
%throughout the plane, which can give rise to a very rich stationary behavior.
%

We would like to emphasize the choice of a linear system is due solely to mathematical convenience, which allow us to develop theoretical analysis with many analytic results. Nevertheless, it is important to bear in mind that, for nonlinear problems, interesting physical behavior can give arise. Particularly, the unusual phenomenon of noise enhancement stability, as discussed by Spagnolo and collaborators \cite{Spagnolo1996,Spagnolo2009,Spagnolo2015,Spagnolo2017}, where nonlinearity and noisy effects contributes to promote enhanced-stability of mean lifetime of metastable and stable states.
Although our model is treated by considering harmonic force field, we find interesting physical results related to steady-states, specially heat flux and entropy generation. 

The time evolution of the system is formulated in terms of an overdamped Brownian dynamics 
in the presence of Langevin forces $\xi_1$ and $\xi_2$ and initial conditions
\begin{equation}
 \begin{split}
  x_i\left( 0 \right) &= 0, \quad \dot{x}_i\left( 0 \right) = 0, \quad i=1,2.
 \end{split} 
\end{equation}
The equation of motion for $x_1$ is given by
%
%\begin{widetext}
\begin{equation} \label{LanSys1}
  \gamma_1 \dot{x}_1\left( t \right) =  -k\,x_1\left( t \right) -k\,u\,x_2\left( t \right) +  {\xi}_1\left( t \right),
\end{equation}
where $\xi_1$ is a Gaussian white noise with cumulants
\begin{equation} \label{NoiseCumu1}
 \begin{split}
  \left<\xi_1(t)\right>_c &= 0, \\
  \left<\xi_1(t)\,\xi_1(t^{\prime})\right>_c &= 2\,\gamma_1\,T_1\,\delta(t-t^{\prime}), \\
 \end{split}
\end{equation}
with temperature $T_1$ and friction coefficient  $\gamma_1$. 
The degree of freedom $x_2$ evolves according to the equation of motion
\begin{equation} \label{LanSys2}
 \begin{split}
  \int_{0}^{t}dt^{\prime}K\left(t - t^{\prime} \right)\, \dot{x}_2(t^{\prime})  =  -k\,x_2\left( t \right) 
  -k\,u\,x_1\left(t \right) + {\xi}_2\left(t \right),
 \end{split}
\end{equation}
%\end{widetext}
%
where $\xi_2$ is a Gaussian colored noise,
\begin{equation} \label{NoiseCumu2}
 \begin{split}
 \left<\xi_2(t)\right>_c &= 0, \\
 \left<\xi_2(t)\,\xi_2(t^{\prime})\right>_c &= \frac{\gamma_2T_2}{\tau}\exp\left( -\frac{|t-t^{\prime}|}{\tau} \right), \\
 \end{split}
\end{equation}
with temperature $T_2$, friction $\gamma_2$ and persistence time-scale $\tau$.
In addition, the second cumulant \eqref{NoiseCumu2} is related to the memory 
kernel by the usual expression
\begin{equation} \label{memorykernel}
 K\left(t - t^{\prime} \right) = \frac{ \left<\xi_2(t)\,\xi_2(t^{\prime})\right>_c  }{T_{2}}.
\end{equation}
If one considers inertial effects, the memory kernel presence still leads to an equilibrium 
steady-state behavior, with a BG statistics, since \eqref{memorykernel} is in 
agreement with the fluctuation-dissipation relation \cite{Kubo1966,Zwanzig}. However, for an overdamped 
Brownian particle, the lack of inertial time-scale and the presence of memory promote a
non-equilibrium stationary distribution which presents an effective local temperature different 
from that associated with thermal bath \cite{NascimentoMorgado2019}.

The coupled stochastic differential equations \eqref{LanSys1} and \eqref{LanSys2} can be rewritten appropriately 
by means of the Laplace-Fourier integral representation, 
\begin{equation}
\tilde{x} \left( s \right) = \int_{0}^{\infty}dt e^{-st}x\left( t \right).
\end{equation}
As a results, the Langevin dynamics reads
%
%\begin{widetext}
\begin{equation} \label{LaplaveLanSys1}
 \begin{split}
    \tilde{x}_1 \left( s \right) = \frac{1}{r\left(s \right)} \Big\{ \left[ {\gamma_2}\,s+k \left( 1+\tau\,s \right)  
     \right] {\tilde{\xi}_1\left( s \right)} - ku \left( 1+\tau\,s \right) {\tilde{\xi}_2\left( s \right) } \Big\},
 \end{split}
\end{equation}      
\begin{equation} \label{LaplaveLanSys2}
 \begin{split}
  \tilde{x}_2\left( s \right) &= \frac{ 1+\tau\,s  }{r\left(s \right)} \left[  \left( {\gamma_1}\,s+k \right) 
    {\tilde{\xi}_2\left( s \right) }-{ ku\, \tilde{\xi}_1 \left( s \right)} \right],
 \end{split}
\end{equation}
%\end{widetext}
%
where
\begin{equation} \label{Requa}
 \begin{split}
     r\left( s \right) &= as^{2}+bs+c = a\left(s-\lambda_{1}\right)\left(s-\lambda_{2}\right),
 \end{split}
\end{equation}
is a quadratic equation with roots $\lambda_{1,2}$ and coefficients
\begin{equation} 
   a = \gamma_1(\gamma_2+k\tau),
\end{equation}
\begin{equation}
   b = k \Big[ \gamma_1 + \gamma_2 + \left( 1-{u}^{2} \right)k\tau\, \Big], 
\end{equation}
\begin{equation}
   c = {k}^{2} \left( 1-{u}^{2} \right).
\end{equation}
Due to the nature of physical parameters in $a,b$ and $c$, it is straightforward to show that $\lambda_{1,2}$ 
assume negative real values, whenever $u^2<1$. In particular, $c$ is related to the stability of the harmonic potential, 
which is well-defined for $u^{2}<1$. 

One can notice from \eqref{LaplaveLanSys1} and \eqref{LaplaveLanSys2} that all cumulants associated 
with the time evolution of the system are straightforwardly obtained in terms of the noise cumulants. 
This is possible due to the kind of potential considered, which allows us to perform all calculations 
analytically. In order to continue our analysis, we should also calculate the Laplace transformation 
of non-zero noise cumulants \eqref{NoiseCumu1} and \eqref{NoiseCumu2}, which gives us
\begin{equation} \label{LaplaceNoiseCumu1}
 \left<\tilde{\xi}_1(s_1)\,\tilde{\xi}_1(s_2)\right>_c = \frac{ 2\,\gamma_1\,T_1}{s_1+s_2},
\end{equation} 
\begin{equation} \label{LaplaceNoiseCumu2}
\left<\tilde{\xi}_2(s_1)\,\tilde{\xi}_2(s_2)\right>_c = {\frac { \Big[ 2+\left( s_1 + s_2\right) \tau \Big] \gamma_{2}\,T_{2}}
 { \left( s_{{1}}+s_{{2}} \right)  \left( 1+s_{{1}}\tau \right) 
 \left( 1 + s_{{2}}\tau \right) }}.
\end{equation}
The solutions of the Langevin equations combined with noise properties allows us to determine all dynamical aspects 
of the system, specially the physical behavior of stationary states.

\section{Stationary probability function} \label{StationaryProb}

From the mathematical perspective, the usual route to calculating the probability distribution associated with a stochastic system 
is by means of a Master Equation (ME) related to the problem \cite{livro_vankampen,Gardiner,Zwanzig,TomeOliveira2015}. For the case of 
Langevin forces described by Gaussian white noises, it is possible to write a 
Fokker-Plank equation, i.e., a continuous example of a ME, and to determine the time-dependent distribution function as well as the stationary behavior 
\cite{Gardiner,Tome2006}. Nevertheless, an alternative method useful for dealing with generalized Langevin 
forces is the time-averaging treatments, where the probability density is calculated by solving the evolution of 
all moments or cumulants \cite{SoaresMorgado2006,SoaresMorgado2008,NascimentoMorgado2019,MorgadoQueiroz2014}. 
These techniques are very appropriate to study systems with complicated kinds of noises, 
such as the white shot noise, or Poisson process \cite{MorgadoQueiroz2016}, and the dichotomous noise 
(telegraph process), which is also a colored-like noise \cite{MedeirosQueiros2015,Queiros2016}. This method yields exact results when potentials are harmonic.

We start by writing the instantaneous distribution function as usual, 
\begin{equation} \label{InstProb}
 \begin{split}
  P( \mathbf{x} ,t) = \left<\delta(  \mathbf{x} - \mathbf{x}(t))\right> 
   = \mathlarger{\int}\frac{d^{2}\mathbf{q}}{4\pi^2}\exp\left( i\mathbf{q} \cdot \mathbf{x} \right)
    G\left( \mathbf{q},t\right),
 \end{split}
\end{equation}
where
\begin{equation}
 \mathbf{q} = 
 \begin{pmatrix}
  q_1 \\
  q_2
 \end{pmatrix},
\end{equation}
and
%\
\begin{equation} \label{InsGen}
 G\left( \mathbf{q},t\right) = \left< \exp\left[ -i\mathbf{q} 
 \cdot \mathbf{x} \left( t \right) \right] \right>,
\end{equation}
is the characteristic function associated with the joint probability density \eqref{InstProb}. 
Notice that both noises present Gaussian structure, which implies that all the moments obtained 
from \eqref{InsGen} can be written in terms of the first and second moments. However, it is more feasible to 
characterize the probability distribution through a cumulant generating function, which depends 
only on the second cumulant for the kind of system we are dealing with. Then, one may write
\begin{equation} \label{CummuFunc}
 \begin{split}
  \ln G( \mathbf{q},t) &= -\frac{1}{2} \mathbf{q} \cdot 
  \begin{pmatrix}
   \mathcal{I}_{11} & \mathcal{I}_{12} \\
   \mathcal{I}_{12} & \mathcal{I}_{22} \\
  \end{pmatrix}
   \cdot \mathbf{q} , 
 \end{split}
\end{equation}
where
%
%\begin{widetext}
\begin{equation} \label{IntCumu}
 \begin{split}
 \mathcal{I}_{ij} = \lim_{\epsilon \to 0} \frac{1}{4\pi^2} \mathlarger{\int \int}  dq_1\, dq_2 e^{\left(iq_1+iq_2+2\epsilon\right)t} 
  \left< \tilde{x}_i\left(iq_1+\epsilon\right)\tilde{x}_j\left(iq_2+\epsilon\right)\right>_c,
  \end{split}
\end{equation}
%\end{widetext}
%
are integrals (with $i,j=1,2$) that account for time evolution contributions of the 
system. In fact, these integrals are the frequency domain representations of the cumulants, as discussed in \ref{AppInt}. Now, we can use the 
Laplace-Fourier form of the Langevin equations \eqref{LaplaveLanSys1} and \eqref{LaplaveLanSys2} combined 
with the expressions for the noise cumulants \eqref{LaplaceNoiseCumu1} and \eqref{LaplaceNoiseCumu2}. 
Then, we have
%
%\begin{widetext}
\begin{equation} \label{Cumu11}
 \begin{split}
  \left< \tilde{x}_1\left(s_1\right)\tilde{x}_1\left(s_2\right)\right>_c &= \Omega  
   \bigg\{ 2\gamma_1T_1\Big[ \gamma_1 s_1 + k\left(  1 + \tau s_1\right) \Big]\Big[ \gamma_2 s_2 + k \left( 1+\tau s_2 \right) \Big] + \bigg. \\ 
   & \qquad \bigg. + \gamma_2T_2\left( ku \right)^{2}\Big[ 2 + \tau \left( s_1 + s_2 \right) \Big] \bigg\},
 \end{split}
\end{equation}
\begin{equation} \label{Cumu12}
 \begin{split}
  \left< \tilde{x}_1\left(s_1\right)\tilde{x}_2\left(s_2\right)\right>_c &= -k\,u\,\Omega  
   \bigg\{ 2\gamma_1T_1 \Big[ \gamma_2 s_1 + k\left( 1 + \tau s_1 \right) \Big]\left( 1 + \tau s_2 \right) + \bigg. \\
   & \qquad \bigg. + \gamma_2T_2\left( k + \gamma_1 s_2 \right)\Big[ 2 + \tau \left( s_1 + s_2 \right) \Big] \bigg\},
\end{split}
\end{equation}
\begin{equation} \label{Cumu22}
 \begin{split}
   \left< \tilde{x}_2\left(s_1\right)\tilde{x}_2\left(s_2\right)\right>_c &= \Omega  
   \bigg\{ 2\gamma_1T_1 \left(ku\right)^{2} \left( 1 + \tau s_1 \right)\left( 1 + \tau s_2 \right) + \bigg. \\ 
   & \qquad \bigg. + \left( k + \gamma_1 s_1 \right)\left( k + \gamma_1 s_2 \right)\Big[ 2 + \tau \left( s_1 + s_2 \right) \Big] \bigg\},
 \end{split}
\end{equation}
%
%\end{widetext}
%the
where
\begin{equation}
 \Omega = \frac{1}{r\left( s_1 \right)r\left( s_2 \right)\left( s_1 + s_2 \right)},
\end{equation}
also depends on variables $s_1$ and $s_2$.

It is possible to calculate an expression for the time-dependent cumulant generating function for the position variables, which exhibits 
many contributions associated with the relaxation process. Although the mathematical structure is quite complicated, it is 
straightforward to notice that there exists basically two important timescales that influence the transients, with one of 
them related with the roots of \eqref{Requa}. Clearly, all those transients are irrelevant for the stationary states, 
which we intend to study. In fact, one can check that nontrivial contributions for the long-term behavior come from  
integrals \eqref{IntCumu} performed around the stationary (thermal) pole, which is obtained by the relation
\begin{equation} \label{thermalpole}
 iq_1 + iq_2 + 2\epsilon =0. 
\end{equation}
As a result, by taking the limit $t \to \infty$, the integrals \eqref{IntCumu} become
\begin{equation} \label{IntCumuX1X1}
 \begin{split}
 \mathcal{I}_{11} &= \lim_{\epsilon \to 0} \frac{1}{2\pi} \mathlarger{\int}  \frac{-2\,dq_1}{ r\left( iq_1 + \epsilon \right)
  r\left( -iq_1 - \epsilon \right) } \Big\{ \left[ \left( \gamma_2 + k\tau \right)^{2}\left( iq_1 + \epsilon \right)^2 -k^{2}  
  \right]\gamma_1\, T_1  - \Big. \\ 
  & \qquad - \Big. \left(k\,u \right)^{2}\gamma_2\,T_2  \Big\},
  \end{split}
\end{equation}
\begin{equation} \label{IntCumuX1X2}
 \begin{split}
 \mathcal{I}_{12} &= \lim_{\epsilon \to 0} \frac{1}{2\pi} \mathlarger{\int}  \frac{2\,k\,u\,dq_1}{ r\left( iq_1 + \epsilon \right)
  r\left( -iq_1 - \epsilon \right) } \Big\{ \left[ \left( \gamma_2 + k\tau \right)\left( iq_1 + \epsilon \right) + k\right] \times \Big. \\ 
    & \qquad \times \Big. \left[ \tau\left( iq_1 + \epsilon \right) -1 \right] \gamma_1\, T_1 -  \left[ \gamma_1\left( iq_1 + \epsilon \right) 
    -k \right]\gamma_2\,T_2  \Big\},
  \end{split}
\end{equation}
\begin{equation} \label{IntCumuX2X2}
 \begin{split}
 \mathcal{I}_{22} &= \lim_{\epsilon \to 0} \frac{1}{2\pi} \mathlarger{\int}  \frac{-2\,dq_1}{ r\left( iq_1 + \epsilon \right)
  r\left( -iq_1 - \epsilon \right) } \Big\{ \left[ \tau^{2}\left( iq_1 + \epsilon \right)^2 -1  \right] \left(k\,u\right)^{2}\gamma_1\, T_1  + \Big. \\
   & \qquad + \Big. \left[  \gamma_{1}^{2}\left( iq_1 + \epsilon \right)^2 -k^{2} \right]\gamma_2\,T_2  \Big\}.
  \end{split}
\end{equation}
Then, performing the remaining integrations, it is possible to write the stationary cumulant generating function as
\begin{equation} \label{CummuFunc1}
 \begin{split}
  \ln G_s( \mathbf{q} ) = -\frac{1}{2}\mathbf{q} \cdot \mathbb{C} \cdot \mathbf{q},
 \end{split}
\end{equation}
where
\begin{equation} \label{Covar}
 \mathbb{C} = 
 \begin{pmatrix}
  \zeta_{11} & \zeta_{12} \\ 
  \zeta_{12} & \zeta_{22} \\
 \end{pmatrix},
\end{equation}
is the covariance matrix which elements are the second cumulants of the distribution, 
%
%
\begin{comment}
%\begin{equation} \label{SetCumu1}
% \begin{split}
%  \beta_{11} &= -\zeta\left( \xi_1\,T_1 + \xi_2\, T_2 \right) , \\
%  \beta_{12} &=  \zeta\left( \xi_3\, T_1 + \xi_4\,  T_2 \right), \\
%  \beta_{22} &=  -\zeta\left( \xi_5\, T_1 + \xi_6\, T_2 \right),
%  \end{split}
%\end{equation}
%
\end{comment}
%
\begin{equation}
  \zeta_{11} = -  \frac{ \left[ {\lambda_1}{\lambda_2}\left( \gamma_2 + k\tau \right)^{2} +{k}^{2} \right] {\gamma_1 T_1 }     
   +  \left(k\,u\right)^{2}\gamma_2 T_2 }{{\lambda_1\lambda_2}\left( {\lambda_1}+{\lambda_2} \right) {a}^{2}} ,
\end{equation}
\begin{equation}
  \zeta_{12} =  \frac{ \left[ {\lambda_1}\,{\lambda_2}{\tau}\left( k+ \tau\,{\gamma_2}\right) +k \right] uk{\gamma_1 T_1} 
   + {k}^{2}u{\gamma_2 T_2} }{{\lambda_1\lambda_2}\left( {\lambda_1}+{\lambda_2} \right) {a}^{2}},
\end{equation}
\begin{equation}
  \zeta_{22} =  -\frac{ \left( {\lambda_1}{\lambda_2}\,{\tau}^{2}+1 \right) \left(k\,u\right)^{2}{\gamma_1}T_1 
   + \left( \,{ {\lambda_1}{\lambda_2}{\gamma_1}}^{2} + {k}^{2}\right) {\gamma_2}T_2 }
   {{\lambda_1\lambda_2}\left( {\lambda_1}+{\lambda_2} \right) {a}^{2}}.
\end{equation}
%
%
\begin{comment}
%
with
%
\begin{equation}
  \xi_1 = \left[ {\lambda_1}{\lambda_2}\left( \gamma_2 + k\tau \right)^{2} +{k}^{2} \right]\gamma_1 ,\quad \xi_2 =  \left(ku\right)^{2}\gamma_2,
\end{equation}
%
\begin{equation}
 \xi_3 = \left[ {\lambda_1}\,{\lambda_2}{\tau}\left( k+ \tau\,{\gamma_2}\right) +k \right] uk\gamma_1, \quad \xi_4 = {k}^{2}u\gamma_2,
\end{equation}
%
\begin{equation}
\begin{split}
 \xi_5 &= \left( {\lambda_1}{\lambda_2}\,{\tau}^{2}+1 \right) \left(ku\right)^{2}\,\gamma_1, \\
  \xi_6 &= \left( {{\lambda_1}{\lambda_2}{\gamma_1}}^{2} + {k}^{2} \right)\gamma_2,
\end{split}
\end{equation}

and
%
\begin{equation} \label{SetCumu11} 
 \zeta = \frac{1}{ {\lambda_1}{\lambda_2}\left( {\lambda_1}+{\lambda_2} \right) {a}^{2} }.
\end{equation} 
%
\end{comment}
%
These cumulants are expressed in terms of the products and sums of the roots of \eqref{Requa}, 
which are related to the coefficients of that equation, in addition to bath temperatures. 
%
\begin{comment}
In particular, the sum ${\lambda_1}+{\lambda_2}$ is negative, since $u^{2}<1$, 
which leads to positive values for $\beta_{11}$ and $\beta_{22}$, as expected. 
\end{comment}
%
Then, we have
%
%
%\begin{widetext}
\begin{equation} \label{SetCumu11}
 \begin{split}
  \zeta_{11} &= \frac{ \left[ c \left( {\gamma_2}+k\tau \right) ^{2 }+a{k}^{2} \right] 
   {\gamma_1}\,{T_1} + a\left(ku\right)^{2}{\gamma_2}\,{T_2} }{abc}, 
  \end{split}
\end{equation}
\begin{equation} \label{SetCumu12}
 \begin{split}
  \zeta_{12} &=  -\frac{ \left[ c\tau\, \left( k+{\gamma_2}\,\tau \right) ^{2} + ak \right] 
   uk {\gamma_1}\,{T_1} + a\,u{k}^{2}\,{\gamma_2}\,{T_2} }{abc},
\end{split}
\end{equation}   
\begin{equation} \label{SetCumu22}
 \begin{split}
  \zeta_{22} &= \frac{ \left( c{\tau}^{2}+a \right) \left(ku\right)^{2}{\gamma_1}\,{T_1} + 
  \left( c{{\gamma_1}}^{2}+a{k}^{2} \right) {\gamma_2}\,{T_2} }{abc}.
  \end{split}
\end{equation}
%\end{widetext}
%
\begin{comment}
where
%
\begin{equation}
 \Upsilon = \frac{1}{abc}.
\end{equation}
\end{comment}

Therefore, the stationary distribution is obtained through the Fourier transform the characteristic function 
that comes from \eqref{CummuFunc1}. Then, we find
\begin{equation} \label{SSDist}
 \begin{split}
  P_{s} \left( \mathbf{x} \right) &= \mathlarger{\int}  \frac{ d^{2}\mathbf{q} }{4\pi^2} 
  \exp\left( -\frac{1}{2}\mathbf{q} \cdot \mathbb{C} \cdot \mathbf{q} + i\mathbf{q}\cdot\mathbf{x}  \right), \\
   &= \frac{1}{2\pi \sqrt{ \textrm{det}\mathbb{C} } } \exp\left( -\frac{1}{2} \mathbf{x} \cdot \mathbb{C}^{-1} \cdot \mathbf{x} \right),
 \end{split}
\end{equation}
where $\textrm{det}\mathbb{C}$ and $\mathbb{C}^{-1}$ are, respectively, the determinant and the inverse of $\mathbb{C}$.
%
\begin{comment}
\begin{equation} \label{InCovar}
 \begin{split}
  \textrm{det}\mathbb{C} &= \beta_{11}\beta_{22}-\beta_{12}^{2}, \\
  \mathbb{C}^{-1} &= \frac{1}{\textrm{det}\mathbb{C}}
  \begin{pmatrix}
   \beta_{22} & - \beta_{12} \\
   -\beta_{12} & \beta_{11} \\
  \end{pmatrix}.
 \end{split}
\end{equation}
\end{comment}
%
Despite its Gaussian character, the general stationary state is not in agreement with the 
BG statistics and, consequently, the system is far-from equilibrium. However, 
for some particular set of model parameters, we can recover the equilibrium properties. 

\subsection{Memoryless limit and different temperatures} \label{T1T2}

For a two-temperature Langevin system subjected to only Gaussian white noises, which corresponds 
 to taking the limit $\tau \to 0$ in \eqref{SetCumu11}-\eqref{SetCumu22}, the cumulants are given by
\begin{equation}
  \zeta_{11} = \frac{ \left( \gamma_1 + \gamma_2 \right)T_1 + \gamma_2u^{2}\left( T_2 - T_1 \right) }
   { k\left( \gamma_1 + \gamma_2 \right) \left( 1-u^{2} \right) },
\end{equation}
\begin{equation}
  \zeta_{12} =  -\frac{ \left( \gamma_1T_1 + \gamma_2T_2 \right)u }{ k\left( \gamma_1 + \gamma_2 \right) \left( 1-u^{2} \right) },
\end{equation}
\begin{equation}
  \zeta_{22} =  \frac{ \left( \gamma_1 + \gamma_2 \right)T_2 + \gamma_1u^{2}\left( T_1 - T_2 \right) }
   { k\left( \gamma_1 + \gamma_2 \right) \left( 1-u^{2} \right) }.
\end{equation}
In the special case of same dissipation mechanisms for both degrees of freedom, $\gamma_1=\gamma_2=\gamma$, 
the covariance matrix is given by
%
%
%\begin{widetext}
\begin{equation} \label{CovarA}
 \mathbb{C} = \frac{1}{2k\left( 1-u^2 \right)}
 \begin{pmatrix}
  2T_1 + \left( T_2-T_1 \right)u^{2} & -\left( T_1 + T_2 \right)u \\ 
  -\left( T_1 + T_2 \right)u & 2T_2 + \left( T_1-T_2 \right)u^{2} \\
 \end{pmatrix}.
\end{equation}
%\end{widetext}
%with
%
%\begin{equation}
% \zeta = \frac{1}{2k\left( 1-u^2 \right)}.
%\end{equation}
%
%
This is in agreement with Dotsenko and collaborators \cite{DotsenkoOshanin2013}, which have shown that 
such a Langevin system presents a non-equilibrium stationary state with spatial-dependent probability 
current. This probability flux leads to a mean rotation velocity, which is associated with a kind to ``symmetry breaking'' rotor, related to interacting degrees of freedom at different effective temperatures. For the case of asymmetric harmonic force fields,  Mancois 
\textit{et al.} \cite{MancoisWilkowski2018} also found, by means of analytic treatments and simulation results,
that different ``temperature axis'' lead to nontrivial current patterns for the steady-state regime. Furthermore, one can find, for asymmetric quadratic potential, a nontrivial average angular velocity that depends on the potential strength $u$ and the difference of temperatures 
$T_2-T_1$ \cite{MancoisWilkowski2018}.

\subsection{Finite memory and same temperatures} \label{Ttau}
Now consider that the baths present the same temperature, $T_1=T_2=T$. As a result, one finds the cumulants
%
%\begin{widetext}
\begin{equation} 
 \zeta_{11} = \frac{ T } { k \left( 1-u^{2} \right) },
\end{equation}
\begin{equation}
 \zeta_{12} =  -{\frac { \left[  \left( {\gamma_1}+{\gamma_2} \right)  \left( { \gamma_2}+k\tau \right) +k\tau\, 
   \left( k+{\gamma_2}\,\tau\right) ^{2} \left( 1-{u}^{2} \right)  \right]u}{ \left[ {\gamma_1}+{\gamma_2} + 
    \tau \left( 1-{u}^{2} \right) \right] \left( {\gamma_2}+k\tau \right)}}\frac{T}{k\left( 1-{u}^{2} \right)},
\end{equation}
\begin{equation}
 \zeta_{22} =  \frac{ \left( \gamma_2 + k \tau u^{2} \right) }
   { \left( \gamma_2 + k \tau \right) }\frac{T}{k\left( 1-{u}^{2} \right)}.
\end{equation}
%\end{widetext}
%
These expressions indicate a non-equilibrium steady-state whenever the memory kernel time-scale is finite, which is an artifact of the overdamped approximation~\cite{NascimentoMorgado2019}. It can be seen that an harmonic 
coupling between the two degrees of freedom is not enough to favour the BG distribution.  Notice that we have just one bath coupled to
each degree of freedom. For a one-dimensional, overdamped, Langevin 
system with colored and white noises, Nascimento and Morgado \cite{NascimentoMorgado2019} have 
shown that it is possible to recover the BG distribution if baths present the 
same temperatures. The present result suggests that dynamical aspects of other baths may 
effectively regularize the inertial contributions on the evolution of massless LD. 
It should be interesting to check whether the inclusion of additional baths may promote 
equilibration for higher dimensional overdamped Brownian dynamics.

In next section we develop investigations of how time-correlated noise may affect the energetic fluxes and 
the entropy generation of the system.
\section{Heat flux and entropy production} \label{HeatEntropy}
The balance equation for the entropy evolution can be written as
\begin{equation} \label{entropuevol}
 \frac{dS}{dt} = \sigma - \phi,
\end{equation}
where $\phi$ is the entropy flux, due to interactions with environment, and 
$\sigma$ is the entropy production inside the system \cite{Prigogine1968,GrootMazur,Callen1985,TomeOliveira2015,Tome2006}. 
The second law of thermodynamics, or minimum entropy production principle, states that $\sigma$ is a non-negative quantity. 
For steady-states, the total entropy $S$ is a stationary quantity, and the entropy generated inside the system should be 
compensated by the entropy flow coming from outside. When the system undergoes reversible processes, $\sigma=\phi=0$, which 
is the typical situation found in equilibrium thermodynamics. On the other hand, a continuous generation of entropy leads 
to stationary states characterized by irreversibility and non-equilibrium distributions.

For our Langevin model, we can determine the stationary behavior of the entropy production by calculating the entropy flow associated with the baths. 
The entropy flow is given by the heat fluxes, which present a temporal evolution consistent with 
the energetic formulation of the Langevin equations. In fact, Sekimoto approach of the stochastic 
thermodynamics\cite{Sekimoto1998,Sekimoto2010,NicolisDecker2017} states that the instantaneous heat fluxes are given by
\begin{equation} \label{heatflux1}
 J_1 \left( t \right) =  \xi_1\left( t \right) \dot{x}_1\left( t \right) - \gamma_1\dot{x}_1^{2}\left(t \right),
\end{equation}
for variable $x_1$, and
\begin{equation} \label{heatflux2}
 J_2 \left( t \right) =  \xi_2\left( t \right) \dot{x}_2\left( t \right) - 
 \int_{0}^{t}dt^{\prime}K\left(t - t^{\prime} \right)\,\dot{x}_2\left( t \right)\, \dot{x}_2(t^{\prime}),
\end{equation}
for degree of freedom $x_2$, which is under the influence of a dissipation memory.
However, due to the harmonic force field character, the Langevin equations \eqref{LanSys1} and \eqref{LanSys2} allows us to rewrite 
\eqref{heatflux1} and \eqref{heatflux2} as 
\begin{equation} \label{heatflux11}
 J_1 \left( t \right) =  k\,x_{1}\left( t \right)\dot{x}_{1}\left( t \right) + 
 k\,u\, x_{2}\left( t \right)\dot{x}_{1}\left( t \right),
\end{equation}
\begin{equation} \label{heatflux22}
 J_2 \left( t \right) =  k\,x_{2}\left( t \right)\dot{x}_{2}\left( t \right) + 
 k\,u\, x_{2}\left( t \right)\dot{x}_{2}\left( t \right).
\end{equation}
The time integral of the average heat flux is the average heat, which gives for $x_1$
\begin{equation} \label{heat1}
 \begin{split}
  \mathcal{Q}_1 \left( t \right) = \int_{0}^{t}dt^{\prime} \left< J_1 \left( t^{\prime} \right) \right>_c 
  = \frac{k}{2}\left< x_{1}^{2} \left( t \right) \right>_c + k\,u\,\psi_t,
 \end{split}
\end{equation}
where
\begin{equation} \label{heatpsi}
  \psi_t = \int_{0}^{t}dt^{\prime} \left<   x_{2}\left( t^{\prime} \right)\dot{x}_{1}\left( t^{\prime} \right) \right>_c.
\end{equation}
%
%
\begin{comment}
\begin{figure}
 \centering
 \includegraphics[scale=0.1]{./figures/2tempA.pdf}
 \caption{Coupled bead-spring analog for a two-dimensional 
  Langevin system in a harmonic trap. Reservoir $1$ is described by a Gaussian white noise with temperature 
  $T_1$. Reservoir $2$, which is at temperature $T_2$, is characterized by a Gaussian colored noise with a dissipation 
  memory kernel. The finite correlation time influences heat fluxes through the degrees of freedom. }
\label{2temp}
\end{figure}
\end{comment}
%
Notice that a relation, similar to \eqref{heat1}, between heat flux and second cumulant is discussed by Ciliberto and collaborators  \cite{CilibertoTanase2013,BerutCiliberto2016}, which studied a two-temperature system under the influence of electric thermal noise.
In fact, earlier theoretical results about energy dissipation and variance is presented by Harada and Sasa 
for a non-equilibrium Langevin system \cite{HaradaSasa2005}. 

Although we determined the energetics related to degree 
of freedom $x_1$,  it is straightforward to perform similar calculations for $x_2$ and perceive that
\begin{equation}
 \left< U \right> = \mathcal{Q}_1 + \mathcal{Q}_2,
\end{equation}
which is the energy conservation form associated with Brownian dynamics. 
We are interested in the properties of the stationary states, for which we have
\begin{equation} \label{energycon}
 \frac{d}{dt} \left< U \right> = \frac{d}{dt}\Big(  \mathcal{Q}_1 + \mathcal{Q}_2 \Big) =0.
\end{equation}
This indicates that we can only focus on the dynamical aspects of just one degree of freedom, say $x_1$.

% \Big[ k\,x_{1}\left( t^{\prime} \right)\dot{x}_{1}\left( t^{\prime} \right) + 
% k\,u\, x_{2}\left( t^{\prime} \right)\dot{x}_{1}\left( t^{\prime} \right) \Big],
%
Notice that the first term in \eqref{heat1} is the second cumulant associated with the degree of freedom $1$, 
which we already determined the for the stationary state. The integral in the second term, given by \eqref{heatpsi}, 
may be calculated by using the Laplace-Fourier formalism, which reads
%
%
%\begin{widetext}
\begin{equation} 
 \begin{split}
  \psi_t = \lim_{\epsilon \to 0}\, \frac{1}{4\pi^2}\mathlarger{\int \int} dq_1\, dq_2  \frac{ e^{\left(iq_1+iq_2+2\epsilon\right)t} -1}
   {\left(iq_1+iq_2+2\epsilon\right)}
   \left( iq_{1} + \epsilon \right) \left< \tilde{x}_1\left(iq_1+\epsilon\right)\tilde{x}_2\left(iq_2+\epsilon\right)\right>_c.
 \end{split}
\end{equation}
%\end{widetext}
%
%
Our main interest here is to determine the stationary properties of the heat flux. The main contributions come from 
the terms that integrate over the residues of the thermal pole \eqref{thermalpole}. Also, we need to derive the second cumulant associated with the coupled degrees of freedom, which we obtained in \eqref{Cumu12}. Now, performing the integral over $q_2$, which can be evaluated without any convergence problems, one obtains
%
%
%\begin{widetext}
\begin{equation} 
 \begin{split}
  \psi_t &= 
   \frac{k\,u\,t}{\pi} \lim_{\epsilon \to 0} \mathlarger{\int} dq_1 \frac{iq_1+\epsilon}
   {\left[ r\left( iq_1 + \epsilon\right)r\left( -iq_1 - \epsilon\right) \right]^{2}} 
    \bigg\{ \Big[ \left( iq_1 + \epsilon \right)\left( k\tau + \gamma_2 \right) + k \Big] \bigg. \times \\ 
   & \qquad \Big. \times \Big[  \left( iq_1 + \epsilon \right)\tau -1 \Big] {\gamma_1}\,T_1 + 
   \Big[ \left(iq_1 + \epsilon \right)\gamma_1 - k \Big] {\gamma_2}\,T_2 \bigg\}.
 \end{split}
\end{equation}
%\end{widetext}
%

%
\begin{figure}
 \centering
 \includegraphics[scale=0.4]{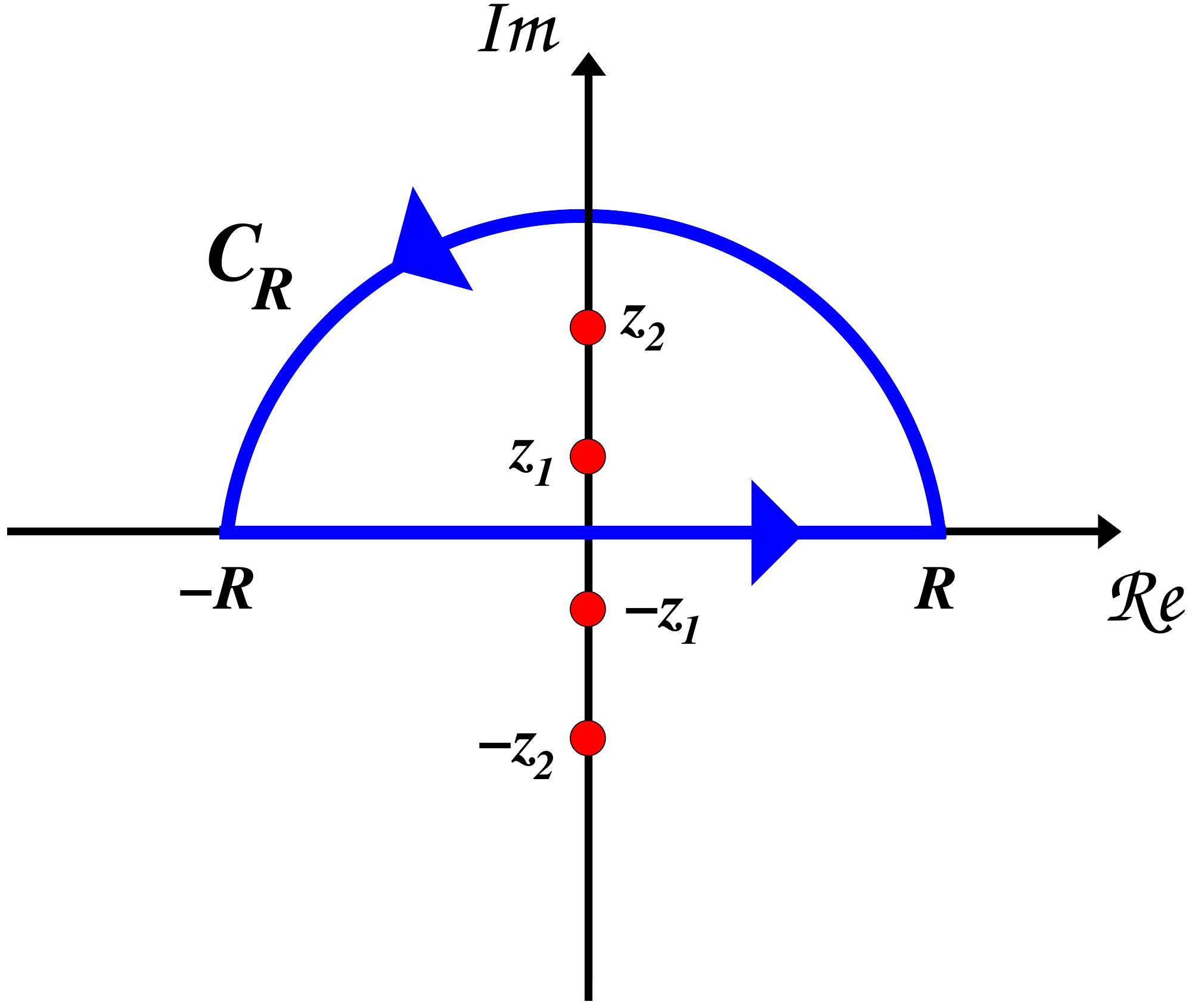}
 \caption{ Typical contour integration associated with steady-state heat fluxes. The poles $\pm z_1$ and $\pm z_2$ are related with 
 roots of equation \eqref{Requa}. Integration along $C_R$ contributes nontrivially for the stochastic energetics.}
\label{contour}
\end{figure}

This  integral over $q_1$ should be calculated through a different approach, since the integrand behaves as $1/q_1$, 
which means that Jordan's lemma is not satisfied for the semi-circular contour part. Nevertheless, it is still 
feasible to evaluate the Cauchy principal value, but we need to consider the nontrivial contributions of the 
semi-circular contour, as schematically represented in Fig.\ref{contour}. This also happens 
if one intends to determine the stochastic energetics of Brownian particles with Poisson white noise 
\cite{MorgadoSoares2011}. Therefore, after performing the integration with appropriate limit procedures, we find
\begin{equation} \label{IntX1X2}
 \psi_t =  -{\frac {ku\, {\gamma_1}{\gamma_2}t}{ \left( {\lambda_1}+
  {\lambda_2} \right) {a}^{2}}}\left( \,{ T_1}-\,{ T_2} \,\right).
\end{equation}
It is worth reinforcing that \eqref{IntX1X2} is valid for large values of $t$.
Using the roots and coefficients of \eqref{Requa}, it is possible to write the long-term 
behavior of \eqref{heat1} as
\begin{equation} \label{heat11}
 \begin{split}
  \mathcal{Q}_1 &= \frac{k\,\zeta_{11}}{2} + \frac{ ku^{2}{\gamma_2}\left( \,{ T_1}-\,{ T_2} \right)t}
  { \left( \gamma_2 + k\tau \right) \Big[ \gamma_1 + \gamma_2 +   k\,\tau\left( 1-u^{2} \right) \Big] },
 \end{split}
\end{equation}
which leads to the stationary heat flux for degree of freedom $x_1$
\begin{equation} \label{StatHeatFlux1}
 \begin{split}
  \mathcal{J}_1=\lim_{t\rightarrow\infty}\left< J_1 \left( t \right)\right>=\frac{d \mathcal{Q}_1}{d t} = {\frac { ku^{2}{\gamma_2}\left( \,{ T_1}-\,{ T_2} 
  \,\right)}{ \left( \gamma_2 + k\tau \right) \Big[ \gamma_1 + \gamma_2 +   k\,\tau\left( 1-u^{2} \right) \Big] }},
 \end{split}
\end{equation}
and from the energetic constraint \eqref{energycon}, we have
\begin{equation} \label{StatHeatFlux2}
 \mathcal{J}_2 = - \mathcal{J}_1.
\end{equation}

Notice that heat flux \eqref{StatHeatFlux1} presents terms of the form $k\,\tau$, which suggests 
an interesting interplay between the oscillator constant parameter $k$ and the memory time-scale $\tau$. 
From a mathematical point of view, it is straightforward to perceive that, for finite values of $\tau$, the heat flux tends to zero as $k \to \infty$. This result should be compared to overdamped 
case of two-dimensional Brownian particle with thermal baths described by Gaussian white noises. 
For this case, one case perceive that, by taking the limit $\tau \to 0$ (in fact $k\tau\rightarrow0$), heat flux \eqref{StatHeatFlux1} 
turns out to depend linearly on spring force constant $k$, which gives rise to a divergence behavior as 
$k \to \infty$. We would like to emphasize that these results should be considered with care because some kind of 
motility of the fluid, which is reduced as spring constant increases, seems to be important from physical considerations concerning the origin of the memory kernel.  
In fact, as discussed by Sekimoto \cite{Sekimoto2010}, heat flux that diverges linearly with $k$ is an artifact of 
lacking inertial contributions, which acts to regularize the heat conduction.

Finally, we can determine, for the steady-state regime, the entropy production by evaluating the  entropy flow 
associated with heat fluxes \eqref{StatHeatFlux1} and \eqref{StatHeatFlux2}. As a result, we have
\begin{equation} \label{entopyprod}
 \begin{split}
  \sigma_s = \mathcal{J}_1 \left( \frac{1}{T_1} - \frac{1}{T_2} \right)= \,\frac{ \nu  \left(\, T_1 - T_2 \,\right)^{2} }{T_1\,T_2},      
 \end{split}
\end{equation}
where
\begin{equation} \label{Gammaentropy}
  \nu = {\frac { {\gamma_2}\,k\,u^{2} }
   { \left( \gamma_2 + k\,\tau \right) \Big[ \gamma_1 + \gamma_2 +   k\,\tau\left( 1-u^{2} \right) \Big] }}
\end{equation}
is a parameter with dimension of inverse time.
The entropy production \eqref{entopyprod} tends to zero when bath temperatures are the same. Nevertheless, 
for finite memory time-scale $\tau$, the stationary distribution is out-of-equilibrium, as discussed in 
\ref{Ttau}. We like to emphasize this peculiar non-equilibrium situation is not inconsistent with the second 
law of thermodynamics. A system which also presents stationary behavior with zero entropy production is a 
massive Brownian particle with Poisson white noise, according to the stochastic energetics 
analysis of Morgado and Guerreiro \cite{MorgadoGuerreiro2012}. It should be said that a Poisson reservoir (a kind of work reservoir) produces entropy by itself. 

It is worth mentioning the findings about irreversibility and memory discussed by Pluglisi and Villamaina 
\cite{PuglisiVillamaina2009}, which show that an  overdamped Brownian particle may present a zero entropy production for 
finite memory time-scales (associated with effectively non-conservative force fields). Also, Langevin equations describe the evolution of effective dynamics, which means that one should choose with care the appropriate set of degrees of freedom. In fact, as discussed by Villamaina and collaborators 
\cite{VillamainaVulpiani2009}, equilibrium behavior is achieved when effective degrees of freedom  are well characterized, which requires generalized fluctuation-dissipation relations.

Another point is that, according to \eqref{Gammaentropy}, the memory time-scale affects the amount of entropy production. In fact, taking the memoryless limit, $\tau \to 0$, we recover the usual form of entropy generation for  coupled Langevin equations with different thermal baths given by Gaussian white noises 
\cite{Sekimoto1998,Sekimoto2010}. Consequently, the time-scale associated with non-Markovian dissipation leads to a decreasing 
of the entropy production for an overdamped two-temperature LD. Notice that, acoording to Puglisi and Villamaina calculations \cite{PuglisiVillamaina2009}, the memory effects seem to decrease the entropy generated by a Brownian particle, where memory acts through effective non-conservative forces. 
In our coupled Langevin equations, temporal correlations, associated with noise and dissipation in an overdamping approximation can drive the 
system out of equilibrium, with a steady-state entropy production which decays with increasing memory time-scale.

Therefore, we can argue that a nontrivial interplay between dissipation memory kernel, overdamped approximations 
and many-bath couplings may promote interesting non-equilibrium states and irreversibility behavior even for 
very simple linear, harmonically-bonded particle models.

\section{Conclusions} \label{Conclusions}

We study the role of memory on overdamped, two-dimensional, Brownian dynamics at different temperatures.
The system is described by two coupled degrees of freedom, interacting via harmonic potential, and the baths 
are characterized by stochastic forces represented by Gaussian white and colored noises, and a memory kernel
related to dissipation. We determine analytically, through time-averaging treatments, the long-term probability 
function associated with degrees of freedom, the stationary states are given by a Gaussian distribution, which may 
lead to BG statistics for some models parameters. Nevertheless, for finite memory and same temperatures, the system presents a steady-state of non-equilibrium.

We investigate the stochastic energetics of the model by calculating the heat fluxes associated with each degree of 
freedom, and the stationary behavior of the entropy generation is analyzed. Memory leads to an heat exchange that exhibits a non-linear dependence with spring force constant $k$, which is in contrast to the linear behavior found as temporal correlations (associated with colored noise) $\tau$ goes to zero. In fact, we find that heat flux decays with $k$ for finite $\tau$, which suggests a very different behavior for high stiffness limit when compared with memoryless case ($\tau \to 0$). Also, memory affects irreversibility associated with steady-states, which present a decaying entropy generation 
with the noise temporal correlations. In particular, we show that, for finite memory time-scale and equal bath temperatures, the system exhibits a non-Gibbsian stationary state with null entropy production.

It seems reasonable to consider further investigations with additional baths, of non-Gaussian as well non-Markovian type, 
in order to understand the role of memory on irreversibility aspects of massive and overdamped systems, specially fluctuation relations and entropy production \cite{Kurchan1998,MarconiVulpiani2008,BerutCiliberto2016A}. Also, it can be interesting to study the interplay between dissipation memory, multiple baths action with distinct effective temperatures, and nonlinear potentials (typical for complex systems), since such as noisy systems present unusual phenomena of stability enhancement induced by stochastic forces 
\cite{Spagnolo1996,Spagnolo2015,Spagnolo2017}. 

\section*{Acknowledgement}
This work is supported by the Brazilian agencies CAPES and CNPq.

\appendix
\section{Laplace-Fourier integral representation} \label{AppInt}

In this is paper we make use of integral representations for the cumulants, which may be written as
\begin{equation} \label{IntCumuAA}
 \left< x^{n}\left( t\right) \right>_c = \left[ \prod_{j=1}^{n} \int d t_{j}\, \delta\left( t - t_j \right) \right]
  \left< x\left( t_1 \right) \cdots x\left( t_n \right) \right>_c,
\end{equation}
where $\delta\left( t - t_j \right)$ is a Dirac delta function. Now, consider the identity 
\begin{equation}
 \delta\left( t - t_j \right) = \lim_{\epsilon \to 0} \frac{1}{2\pi} \int dq_j\, e^{\left( iq_j + \epsilon \right)\left( t - t_j \right)}.
\end{equation}
This Fourier integral for delta function allows to rewrite \eqref{IntCumuAA} as 
\begin{equation} \label{A1}
 \left< x^{n}\left( t\right) \right>_c = \lim_{\epsilon \to 0} \left( \frac{1}{2\pi} \right)^{n} 
  \left[ \prod_{j=1}^{n} \int d q_{j}\, e^{\left( iq_j + \epsilon \right)t } \right]
  \left< \tilde{x}\left( iq_1 + \epsilon \right) \cdots \tilde{x}\left( iq_n + \epsilon \right) \right>_c,
\end{equation}
where
\begin{equation}
 \tilde{x}\left( s \right) = \int_{0}^{\infty} dt\, e^{-st} x\left( t \right),
\end{equation}
is the Laplace transform of $x\left( t \right)$.
Clearly, the same approach is valid for dealing with the temporal evolution of moments.

%\bibliographystyle{apsrev4-1}
%\bibliographystyle{unsrt}
%\bibliography{REF}

\end{document}